\documentstyle[floats,twocolumn,prb,aps,epsf]{revtex}
\begin{document}
\draft
\preprint{}
%%%%%%%%%%%%%%%%%%%%%%%%%%%%%%%%%%%%%%%%%%%%%%%%%%%%%%%%%
\title{Magnetic Properties for the  One-Dimensional 
Multicomponent Spin-Gap System
}
%%%%%%%%%%%%%%%%%%%%%%%%%%%%%%%%%%%%%%%%%%%%%%%%%%%%%%
\author{Akira Kawaguchi, Tatsuya Fujii and Norio Kawakami} 
\address{Department of Applied Physics,
Osaka University, Suita, Osaka 565-0871, Japan} 
\date{\today}
\maketitle
%----------------------------------------------------------------------
%                              Abstract
%----------------------------------------------------------------------
\begin{abstract}
Magnetic properties for the one-dimensional multicomponent
quantum spin system with the excitation gap are studied based on the 
integrable spin model introduced by Bariev {\it et al}. 
By exactly computing  
the magnetization, we show how the characteristic
structure with plateaus and cusps appears in the magnetization process.
To study low-energy dynamics of the system, we apply the finite-size 
scaling analysis to the excitation spectrum, and thereby evaluate the 
power-law exponent as well as the enhancement factor 
 for the low-temperature NMR relaxation rate $1/T_1$.  We discuss 
the critical properties of $1/T_1$ around  plateaus and 
cusps in the magnetization curve.
\end{abstract}

\pacs{PACS numbers: 75.10.Jm, 75.40.Gb}

%%%%%%%%%%%%%%%%%%%%%%%%%%%%%%%%%%%%%%%%%%%%%%%%%%%%%%%%%%%%%%%%%%%
%%%%%%%%% 4.1 Introduction  　　　　  %%%%%%%%%%%%%%%%%%%%%%%%%%%%%
%%%%%%%%%%%%%%%%%%%%%%%%%%%%%%%%%%%%%%%%%%%%%%%%%%%%%%%%%%%%%%%%%%%

\section{Introduction}

Quantum phase transitions of one-dimensional (1D) spin systems
in a magnetic field have attracted much interest recently. 
One of the most remarkable phenomena is the plateau formation 
in the magnetization process,\cite{Hida,Okamoto,Oshikawa,Exp1,Exp2} 
which is caused by a
field-induced spin gap.  Also, the closure of the spin gap in the 
Haldane system and the ladder system by the magnetic field has 
been providing an interesting subject,
\cite{mHaldane,mLadder}
which shares common important physics with the 
plateau formation. 

Another hot topic in 1D spin systems has been concerned with
the multicomponent version of the spin model or the spin-orbital model, 
since it was recognized that the orbital degrees of freedom 
give rise to a variety of interesting phenomena.
Experimentally, 1D correlated electron systems at quarter-filling,
such as ${\rm Na_2Ti_2Sb_2O}$\cite{Axtell} and ${\rm NaV_2O_5}$\cite{Isobe} 
have been studied extensively, 
for which the orbital degrees of freedom play 
an important role.\cite{Fukuyama,Fulde,Ohta}
In this context, the 1D SU(4) massless spin model has been studied  
numerically\cite{ueda,troy} and analytically
\cite{Sutherland,Affleck,itakura,fujii,zhang} as the simplest 
spin-orbital model, and this type of analysis has 
been extended to a perturbed SU(4) model to study the gap formation.
\cite{Pati,Azaria1,Yamashita,itoi,tsukamoto} 

According to the above studies, it may be interesting 
to investigate how the orbital degeneracy affects field-induced quantum 
phase transitions such as the plateau formation in 1D spin systems.
To address this problem, in this paper, we investigate 
 static and dynamical properties for 
the 1D multicomponent spin-gap system in a magnetic field.
For this purpose, we employ the 1D multicomponent anisotropic
Heisenberg model, which was exactly solved by Bariev {\it et al}.\cite{Bariev}
In particular, we focus our attention on the critical regions
where the spin-gap disappears, or the magnetization plateau (or cusp) is 
generated in the presence of a magnetic field. To study 
the low-energy dynamics of the system, we further
discuss the NMR relaxation rate $1/T_1$ around 
the field-induced quantum phase transitions at low temperatures.
\cite{schulz,sachdev1,chitra,AK,takigawa,chaboussant,Goto} 

This paper is organized as follows.  In Sect.II, we exactly calculate 
the magnetization curve for the 1D anisotropic Heisenberg 
model,\cite{Bariev} and determine the magnetic phase diagram.  
We clarify how the plateaus and cusps are generated in the magnetization 
curve. In Sect.III, by exploiting  finite-size scaling techniques\cite{Cardy} 
we discuss the low-temperature behavior of the 
NMR relaxation rate $1/T_{1}$
with the emphasis on the critical properties 
 around  plateaus and cusps.  It is shown
 that the enhancement factor plays a significant role for $1/T_1$
around such critical regions. A brief summary is given in Sect.IV. 

%%%%%%%%%%%%%%%%%%%%%%%%%%%%%%%%%%%%%%%%%%%%%%%%%%%%%%%%%%%%%%%%%%%
%%%%%%%%% 4.2　magnetization　　　　  %%%%%%%%%%%%%%%%%%%%%%%%%%%%%
%%%%%%%%%%%%%%%%%%%%%%%%%%%%%%%%%%%%%%%%%%%%%%%%%%%%%%%%%%%%%%%%%%%

\section{ Plateaus and Cusps in the Magnetization}

In this section, we study the magnetization process
for the anisotropic Heisenberg model with a two-band 
structure in 1D, which is defined by the 
deformed version of the SU(4) Heisenberg
model,\cite{Bariev}
%%%%%%%%%%%%%%%%%%%%%%%%%%%%%%%%%%%%%%%%%%%%%%%%%%%%%%%%%%%%%%%%%%%%%%%%%%
\begin{eqnarray}
{\cal H}_{0}&=&J
      \sum_{i}\sum_{m\neq m'}\{
         c_{i m}^{\dagger}c_{i m'}c_{i+1 m'}^{\dagger}c_{i+1 m}
   \cr
   &&
            -\exp[{\rm sign}(m'-m)\Phi]n_{i m}n_{i+1 m'}
               \},
\label{XXZ}
\end{eqnarray}
%%%%%%%%%%%%%%%%%%%%%%%%%%%%%%%%%%%%%%%%%%%%%%%%%%%%%%%%%%%%%%%%%%%%%%%%%%
where it is prohibited that more than one 
electron occupies each site. Here, 
$c_{i m}^{\dagger}$ is the creation operator for electrons 
with  spin ($\uparrow,\downarrow$) and orbital ($1,2$) degrees 
of freedom. 
We shall use  the indices $m=1,2,3,4$  for these 
four states so that the corresponding 
number of electrons $N_m$ should satisfy, $N_1\geq N_2\geq N_3\geq N_4$
 in a magnetic field.
In the isotropic case ($\Phi=0$), 
the Hamiltonian (\ref{XXZ}) is reduced to
the SU(4) massless Heisenberg model,\cite{Sutherland}
whereas for any other cases
it is massive for all excitations at zero magnetic field.\cite{Bariev}
Although the above integrable model is rather special
in its appearance, we think that its characteristic behavior
in a magnetic field, such as the 
formation of plateaus and cusps, should capture some
generic properties expected for multicomponent spin-gap models in 1D.

The exact solution of the model (\ref{XXZ}) with periodic boundary 
conditions was obtained by Bariev {\it et al}.\cite{Bariev}
The Bethe equations obtained  for rapidities
 $\lambda_{j}^{(\alpha)}$ ($\alpha=1,2,3$) read\cite{Bariev,Sato},
%%%%%%%%%%%%%%%%%%%%%%%%%%%%%%%%%%%%%%%%%%%%%%%%%%%%%%%%%%%%%%%%%%%%%%%%%%%%
\begin{eqnarray}
2\pi I_{j}^{\alpha}
         -N\Theta_{1}(\lambda_{j}^{\alpha})\delta_{\alpha 1}
    =&&
          \sum_{s=\pm 1}\sum_{i=1}^{M_{\alpha+s}}
                 \Theta_{1}(\lambda_{j}^{\alpha}-\lambda_{i}^{\alpha+s})
    \cr 
    &&
         -\sum_{i}^{M_{\alpha}}
                 \Theta_{2}(\lambda_{j}^{\alpha}-\lambda_{i}^{\alpha}),
\label{XXZ-Bethe}
\end{eqnarray}
%%%%%%%%%%%%%%%%%%%%%%%%%%%%%%%%%%%%%%%%%%%%%%%%%%%%%%%%%%%%%%%%%%%%%%%%%%%%
with $M_{\alpha}=\sum_{m=\alpha}^{4}N_m$ ($M_{0}=M_{4}=0$),
where $\Theta_{n}(\lambda)=
2\tan^{-1}(\cosh n\Phi\tan\frac{\lambda}{2})$.
The quantum number $I_{j}^{(\alpha)}$ which specifies 
elementary  excitations is subject to the constraints,
%%%%%%%%%%%%%%%%%%%%%%%%%%%%%%%%%%%%%%%%%%%%%%%%%%%%%%%%%%%%%%%%%%%%%%%%%%%%
\begin{eqnarray}
    I_{j}^{(1)}&=&\frac{1}{2}(M_1+M_2) \;\;{\rm mod} \;\; 1,
\nonumber \\
    I_{j}^{(2)}&=&-\frac{1}{2}(M_1-M_2+M_3) \;\;{\rm mod} \;\; 1,
\nonumber \\
    I_{j}^{(3)}&=&\frac{1}{2}(M_2-M_3) \;\;{\rm mod} \;\; 1.
\end{eqnarray}
%%%%%%%%%%%%%%%%%%%%%%%%%%%%%%%%%%%%%%%%%%%%%%%%%%%%%%%%%%%%%%%%%%%%%%%%%
 In the  thermodynamic limit at zero temperature,
 the algebraic equations (\ref{XXZ-Bethe}) are converted to 
the linear integral equations for the distribution functions 
of rapidities $\sigma_{\alpha}$ and 
for the dressed energies $\varepsilon_{\alpha}$, 
respectively,\cite{Bariev,Sato}
%%%%%%%%%%%%%%%%%%%%%%%%%%%%%%%%%%%%%%%%%%%%%%%%%%%%%%%%%%%%%%%%%%%%%%%%%%%%
\begin{eqnarray}
\varepsilon_{\alpha}(\lambda_{\alpha})
         =&&
                  \varepsilon_{\alpha}^{0}(\lambda_{\alpha})
       \cr
       &&
          +\sum_{\gamma=1}^{3}
          \int_{-\lambda_{\gamma}^{0}}
               ^{+\lambda_{\gamma}^{0}}
                 \frac{{\rm d}\lambda'_{\gamma}}{2\pi}
               K_{\alpha\gamma}(\lambda_{\alpha}-\lambda'_{\gamma})
                \varepsilon_{\gamma}(\lambda'_{\gamma}),
\label{XXZ-energy}
\\
\sigma_{\alpha}(\lambda_{\alpha}) 
        =&&
                  \sigma_{\alpha}^{0}(\lambda_{\alpha})
        \cr
        &&
          +
            \sum_{\gamma=1}^{3}\int_{-\lambda_{\gamma}^{0}}
               ^{+\lambda_{\gamma}^{0}}
               \frac{{\rm d}\lambda'_{\gamma}}{2\pi}
               K_{\alpha\gamma}(\lambda_{\alpha}-\lambda'_{\gamma})
                \sigma_{\gamma}(\lambda'_{\gamma}),
\label{XXZ-sigma}
\end{eqnarray}
%%%%%%%%%%%%%%%%%%%%%%%%%%%%%%%%%%%%%%%%%%%%%%%%%%%%%%%%%%%%%%%%%%%%%%%%%%%%
where $K_{\alpha\gamma}=-K_{2}\delta_{\alpha,\gamma}
+K_{1}(\delta_{\alpha,\gamma+1}+\delta_{\alpha+1,\gamma})$,
and $K_{n}(\lambda)=\sinh(n\Phi)/[\cosh(n\Phi)-\cos\lambda]$.
Here, the bare distribution function is given by 
%%%%%%%%%%%%%%%%%%%%%%%%%%%%%%%%%
\begin {eqnarray}
\sigma_\alpha^0(\lambda_1)=\frac{1}{2\pi}K_1(\lambda_1) \delta_{\alpha 1},
\end{eqnarray}
%%%%%%%%%%%%%%%%%%%%%%%%%%%%%%%%%%
and $\varepsilon_{\alpha}^{0}$ 
will be explicitly given separately for each case.
The cut-off parameters ("Fermi level" for elementary
excitations) $\lambda_{\alpha}^{0} (0\leq\lambda_{\alpha}^{0}\leq\pi)$
are determined by minimizing the free energy for a given
magnetic field, resulting in the condition,\cite{Bariev,Sato}
 $\varepsilon_{\alpha}(\pm\lambda_{\alpha}^{0})=0$.
For zero magnetic field, they are reduced to
$\lambda_{1}^{0}=\lambda_{2}^{0}=\lambda_{3}^{0}=\pi$. 
Since the dressed energies satisfy
$\varepsilon_{1}(\pm\lambda_{1}^{0}=\pi)
=\varepsilon_{2}(\pm\lambda_{2}^{0}=\pi)
=\varepsilon_{3}(\pm\lambda_{3}^{0}=\pi)<0$, 
the system has three kinds of 
spin-gap excitations, which are degenerate at zero magnetic field.

As mentioned above, one of the most remarkable phenomena in 1D 
spin systems is the plateau formation in the 
magnetization,\cite{Hida,Okamoto,Oshikawa,Exp1,Exp2}  which is regarded as 
a quantum phase transition driven by the magnetic field.
Also, the middle-field cusp singularity\cite{Okunishi} is another
typical example of the field-induced quantum phase transition.
In the following, we show that the multicomponent model 
(\ref{XXZ}) possesses a rich phase diagram in a magnetic field,
where the magnetization has plateaus and/or 
cusps depending on the anisotropy as well as the
Zeeman splitting. To demonstrate these things, we deal with two 
typical cases for which the energy splitting due to the 
magnetic field appears in different ways.
 
%%%%%%%%%%%%%%%%%%%%%%%%%%%%%%%%%%%%%%%%%%%%%%%%%%%%%%%%%%%%%%%%%%%%%
%%%%%%%%%%%%%%%%%%%% spin-orbit multiplet    　　　%%%%%%%%%%%%%%%%%%
%%%%%%%%%%%%%%%%%%%%%%%%%%%%%%%%%%%%%%%%%%%%%%%%%%%%%%%%%%%%%%%%%%%%%
\subsection{Spin-orbit multiplet case}

We first assume that the Zeeman splitting is given by 
$g\mu_B M_z H$ with $M_z=\pm 1/2, \pm \gamma/2$,
where $\gamma$ ranges from $0$ to $1$.
The corresponding Hamiltonian is 
%%%%%%%%%%%%%%%%%%%%%%%%%%%%%%%%%%%%%%%%%%%%%%%%%%%%%%%%%%%%%%%%%%%%%
\begin{eqnarray}
{\cal H}_{(A)}= -\frac{1}{2}h
             \sum_i [&&
             (n_{i,1/2}-n_{i,-1/2})
     \cr
             &&+\gamma(n_{i,\gamma/2}-n_{i,-\gamma/2})
              ],
\label{spin-orbit}
\end{eqnarray}
%%%%%%%%%%%%%%%%%%%%%%%%%%%%%%%%%%%%%%%%%%%%%%%%%%%%%%%%%%%%%%%%%%%%%%%
where  $n_{i, \alpha}$ denotes the particle number for
electrons with $M_z=\alpha$.
Note that the case of  $\gamma=1/3$ corresponds to the  
Zeeman splitting for the 
spin-orbit multiplet $J_z=\pm 3/2,\pm 1/2$ ($M_z=\frac{1}{3}J_z$),
while for
 $\gamma=1$,  the Zeeman effect acts only on the spin sector,
so that the four-fold multiplet splits into two 
orbitally-degenerate levels in magnetic fields. 
%##The Zeeman term  (\ref{spin-orbit}) may be applied to more
%##generic cases including the above two important ones.
The Zeeman term  (\ref{spin-orbit}) may be applied to more 
generic cases including these two important cases. 
For example, in some Ce compounds such as CeB$_6$ the four-fold 
ground multiplet ($\Gamma_8$ multiplet) is generated in a cubic 
crystalline field under strong spin-orbit coupling. In such cases, 
the value of $\gamma$ is known to depend on the strength of 
the crystalline filed.  We thus regard $\gamma$ as a continuously 
changing parameter in the following discussions. 
We define the magnetization in terms of the distribution functions, 
%%%%%%%%%%%%%%%%%%%%%%%%%%%%%%%%%%%%%%%%%%%%%%%%%%%%%%%%%%%%%%%%%%%%%%%%%%%%
\begin{eqnarray}
&&m_{(A)}=\left\{
         (N_{1/2}-N_{-1/2})
                     +\gamma(N_{\gamma/2}-N_{-\gamma/2})
         \right\}/L
       \nonumber \\
       &&= 
           1-2\gamma\int_{-\lambda_2^0}^{+\lambda_2^0}
                    \sigma_2(\lambda_2){\rm d}\lambda_2
       \cr
       &&
        - (1-\gamma)
          \left\{
                 \int_{-\lambda_1^0}^{+\lambda_1^0}
                      \sigma_1(\lambda_1){\rm d}\lambda_1
                 +\int_{-\lambda_3^0}^{+\lambda_3^0}
                      \sigma_3(\lambda_3){\rm d}\lambda_3
           \right\}, 
  \label{definition-mag-A}         
\end{eqnarray}
%%%%%%%%%%%%%%%%%%%%%%%%%%%%%%%%%%%%%%%%%%%%%%%%%%%%%%%%%%%%%%%%%%%%%%%%%%%%
where we have set the number of electrons as $N_1=N_{1/2}, 
N_2=N_{\gamma/2}, N_3=N_{-\gamma/2}$ and $N_4=N_{-1/2}$, respectively. 
Here $\varepsilon_{\alpha}^{0}$ in (\ref{XXZ-energy}) is given as 
%%%%%%%%%%%%%%%%%%%%%%%%%%%%%%%%%%%%%%%%%%%%%%%%%%%%%%%%%%%%%%%%%%%%%%%%%%%%
%%\begin{eqnarray}
%%    \left(
%%        \begin{array}{c}
%%            \sigma_1^0 \\
%%            \sigma_2^0 \\
%%            \sigma_3^0
%%        \end{array}
%%    \right) 
%%&=&
%%    \left(
%%        \begin{array}{c}
%%            \frac{1}{2\pi}K_1(\lambda_1) \\
%%            0 \\
%%            0
%%        \end{array}
%%    \right)
%%\label{XXZ-sigma0}
%%\\
%%%%%%%%%%%%%%%%%%%%%%%%%%%%%%%%%%%%%%%%%%%%%%%%%%%%%%%%%%
\begin{eqnarray} 
 \left(
        \begin{array}{c}
            \varepsilon_1^0 \\
            \varepsilon_2^0 \\
            \varepsilon_3^0
        \end{array}
    \right) 
=
    \left(
        \begin{array}{c}
            -\sinh \Phi K_1(\lambda_1)+\frac{1}{2}(1-\gamma)h \\
            \gamma h \\
            \frac{1}{2}(1-\gamma)h
        \end{array}
    \right).
\end{eqnarray}
%%%%%%%%%%%%%%%%%%%%%%%%%%%%%%%%%%%%%%%%%%%%%%%%%%%%%%%%%%%%

We have performed the calculation by iterating eq.(\ref{XXZ-energy})
to determine the cutoff parameter $\lambda_{\gamma}^{0}$ 
for a given magnetic field, and then have evaluated the magnetization
by solving eq.(\ref{XXZ-sigma}).
Let us  start with  the spin-orbit level scheme for the 
multiplet specified by the total angular momentum.
%%%%%%%%%%%   FIGURE %%%%%%%%%%%%%%%%%%%%%%%%%%%%%%%%%%%%%%%%
\begin{figure}[tb]
\begin{center}
\vspace{-0.0cm}
\hspace{-0.5cm}
\leavevmode \epsfxsize=125mm 
\epsffile{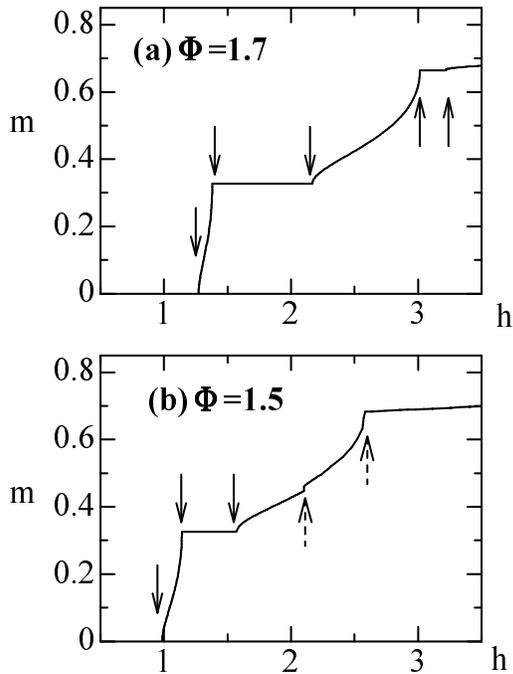}
\vspace{-7.5cm}
\end{center}
\caption{The magnetization curve for (a) $\Phi=1.7$ and (b)$\Phi=1.5$: 
We denote the quantum phase transition points by the arrows. 
In particular, the positions of cusps are shown by 
the arrows with broken line. The saturated magnetization is $m=1$. 
}
\label{mag}
\end{figure}
%%%%%%%%%%%%%%%%%%%%%%%%%%%%%%%%%%%%%%%%%%%%%%%%%%%%%%%%%%%%%%%%%%%
In Figs. \ref{mag}(a) and (b), 
the  magnetization curve for  $\gamma=1/3$ is shown 
 by choosing two different values of 
 the anisotropy, $\Phi=1.7$ and $\Phi=1.5$.
When the magnetic field is increased from zero, 
the magnetization starts to increase at the 
finite critical field with a square-root dependence.
For $\Phi=1.5$, the $m=1/3$ plateau appears,
 which is followed by  two cusps in 
higher magnetic fields.  On the other hand  for $\Phi=1.7$, 
 two plateaus  appear at $m=1/3$ and $m=2/3$, and 
the cusp structures are not generated. 
%##Assuming that the effective spin moment is given by 
%##spin $J=3/2$ multiplet, it is confirmed that the formation of 
%##these plateaus satisfies the criterion for 
%##the fractional quantization introduced  by 
%##M. Oshikawa {\it et al}.\cite{Oshikawa}. 
We note that the formation of plateaus in the present model is 
different from the standard fractional quantization discussed by 
Oshikawa {\it et al}.\cite{Oshikawa}, for which the plateau is 
stabilized by the spin-gap state that has magnetic order commensurate 
with the underlying lattice structure. On the other hand, in the 
present case, the plateau is generated when one of the multicomponent 
spin modes becomes massive, and the corresponding ground state is 
still in a disordered state which is not accompanied by the 
commensurate spin arrangement. 

%%%%%%%%%%%%%%%%%%%%%%%%%%%%%%%%%%%%%%%%%%%%%%%%%%%%%%%%%%%%%%%%%%%
\begin{figure}[h]
\begin{center}
\vspace{-0cm}
\leavevmode \epsfxsize=80mm 
\epsffile{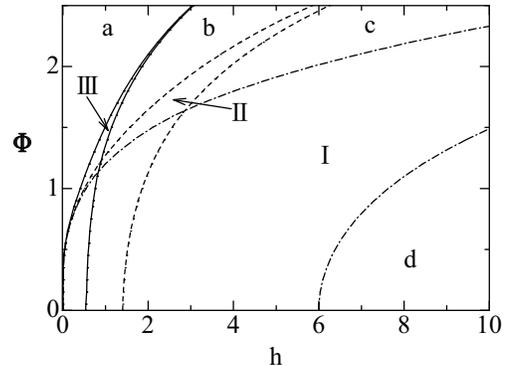}
\vspace{-6.0cm}
\end{center}
\caption{  The $h-\Phi$ phase diagram 
for the model (\protect{\ref{XXZ}}) with (\protect{\ref{spin-orbit}}) 
for $\gamma=1/3$: 
The regions I, II and III denote  the areas which are 
respectively sandwiched  by two dash-dotted lines,
two broken lines, and two solid lines, where  
each elementary excitation,  
$\varepsilon_1$, $\varepsilon_2$, or $\varepsilon_3$ 
is massless. Magnetization plateaus  appear 
in the spin-gap phases outside of the above  three regions:
(a) $m=0$, (b) $m=1/3$, (c) $m=2/3$, (d) $m=1$. 
Magnetization cusps are observed 
at the boundary where two kinds of massless regions overlap.
}
\label{h-phi}
\end{figure}
%%%%%%%%%%%%%%%%%%%%%%%%%%%%%%%%%%%%%%%%%%%%%%%%%%%%%%%%%%%%%%%%%%%

%%%%%%%%%%%%%%%%%%%%%%%%%%%%%%%%%%%%%%%%%%%%%%%%%%%%%%%%%%%%%%%%%%%
\begin{figure}[h]
\begin{center}
\vspace{-0cm}
\leavevmode \epsfxsize=80mm 
\epsffile{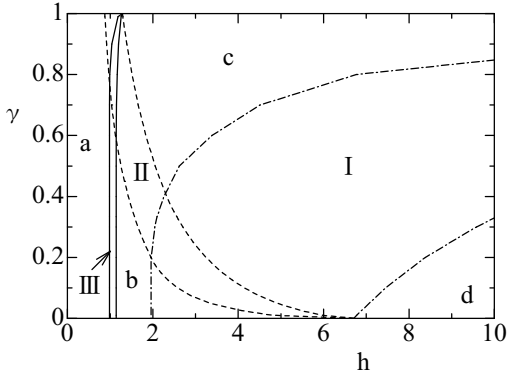}
\vspace{-6cm}
\end{center}
\caption{The $h-\gamma$ phase diagram 
for the model (\protect{\ref{XXZ}}) with (\protect{\ref{spin-orbit}}) 
at $\Phi=1.5$ 
Each region is specified following the way in 
Fig.\protect{\ref{h-phi}}.
}
\label{h-gamma}
\end{figure}
%%%%%%%%%%%%%%%%%%%%%%%%%%%%%%%%%%%%%%%%%%%%%%%%%%%%%%%%%%%%%%%%%%%

In Fig. \ref{h-phi}, the $h-\Phi$ phase diagram is shown 
 for the case of spin-orbit level scheme with  $\gamma=1/3$. 
Note that a massless mode exists in the regions I,  II and III, in which
 the magnetization increases continuously with the field. 
In the area where two of  I, II and III overlap
with each other, there exist two  massless modes. 
The cusp structure appears just at 
the boundary where these massless regions overlap. 
On the other hand, the regions of (a)-(d) outside of the 
massless regions denote the spin-gap phases, where 
the magnetization exhibits plateaus quantized as
(a)$m=0$, (b)$m=1/3$, (c)$m=2/3$, (d)$m=1$, respectively.
Note that the $\Phi=0$ line corresponds to the isotropic 
SU(4) model,\cite{Sutherland}  which is known to have  
only two cusps without plateaus.\cite{Okunishi}
 The plateaus appear at $m=1/3$ and $m=2/3$
for sufficiently large  $\Phi$.

For reference, we show in Fig. \ref{h-gamma} the $h-\gamma$ phase diagram 
for $\Phi=1.5$ to see how the scheme of the Zeeman 
splitting affects the magnetization curve.
Recall that  for the case of $\gamma=1$, the four-fold 
degenerate states split into  two-fold orbitally
degenerate states in a magnetic field,
for which  the plateau-structure is absent. 
If  the value of the parameter $\gamma$ deviates from
$\gamma=1/3,1$,  the plateau appears at $m=(1+\gamma)/2$
 in the  region (c).

%%%%%%%%%%%%%%%%%%%%%%%%%%%%%%%%%%%%%%%%%%%%%%%%%%%%%%%%%%%%%%%%%%%%%%%%%%%%
%%%%%%%%%%%%%%%%%%%%　2.2                    %%%%%%%%%%%%%%%%%%%%%%%%%%%%%
%%%%%%%%%%%%%%%%%%%%　　crystal field　　　　　%%%%%%%%%%%%%%%%%%%%%%%%%%%%%
%%%%%%%%%%%%%%%%%%%%%%%%%%%%%%%%%%%%%%%%%%%%%%%%%%%%%%%%%%%%%%%%%%%%%%%%%%%%

\subsection{Multiplet in a crystal field}

We give another example of the magnetic phase diagram 
by employing  a slightly different 
level scheme for which the orbital-splitting $\Delta$ 
due to the crystal field exists at zero magnetic field, 
and these two distinct levels further
split in the presence of the magnetic field $h$. 
The corresponding Hamiltonian is
%%%%%%%%%%%%%%%%%%%%%%%%%%%%%%%%%%%%%%%%%%%%%%%%%%%%%%%%%%%%%%%%%%%%%%%%%%%%
\begin{eqnarray}
{\cal H}_{(B)}=\sum_i \left[
           -\frac{1}{2}h(n_{i,\uparrow}-n_{i,\downarrow})
             -\Delta(n_{i,1}-n_{i,2})
             \right], 
 \label{crystal}
\end{eqnarray}
%%%%%%%%%%%%%%%%%%%%%%%%%%%%%%%%%%%%%%%%%%%%%%%%%%%%%%%%%%%%%%%%%%%%%%%%%%%%
where $n_{\uparrow}=n_{1\uparrow}+n_{2\uparrow}$ 
and $n_{1}=n_{1\uparrow}+n_{1\downarrow}$. The  calculation 
of the magnetization is performed straightforwardly in the way
outlined above.
For $h\leq 2\Delta$, we set the number of electrons as $N_1=N_{1\uparrow}, 
N_2=N_{1\downarrow}, N_3=N_{2\uparrow}$ and $N_4=N_{2\downarrow}$, 
respectively. 
In this case, $\varepsilon_{\alpha}^{0}$ is given by
%%%%%%%%%%%%%%%%%%%%%%%%%%%%%%%%%%%%%%%%%%%%%%%%%%%%%%%%%%%%%%%%%%%%%%%%%%%%
\begin{eqnarray}
   \left(
        \begin{array}{c}
            \varepsilon_1^0 \\
            \varepsilon_2^0 \\
            \varepsilon_3^0
        \end{array}
    \right) 
=
    \left(
        \begin{array}{c}
            -\sinh \Phi K_1(\lambda_1)+h \\
            -h+2\Delta \\
            h
        \end{array}
    \right).
\end{eqnarray}
%%%%%%%%%%%%%%%%%%%%%%%%%%%%%%%%%%%%%%%%%%%%%%%%%%%%%%%%%%%%%%%%%%%%%%%%%%%%
 On the other hand, for $h\geq 2\Delta$, we set the number
of electrons $N_1=N_{1\uparrow}, 
N_2=N_{2\uparrow}, N_3=N_{1\downarrow}$ and 
$N_4=N_{2\downarrow}$, for which $\varepsilon_{\alpha}^{0}$ reads
%%%%%%%%%%%%%%%%%%%%%%%%%%%%%%%%%%%%%%%%%%%%%%%%%%%%%%%%%%%%%%%%%%%%%%%%%%%%
\begin{eqnarray}
   \left(
        \begin{array}{c}
            \varepsilon_1^0 \\
            \varepsilon_2^0 \\
            \varepsilon_3^0
        \end{array}
    \right) 
=
    \left(
        \begin{array}{c}
            -\sinh \Phi K_1(\lambda_1)+2\Delta \\
            h-2\Delta \\
            2\Delta
        \end{array}
    \right).
\end{eqnarray}
%%%%%%%%%%%%%%%%%%%%%%%%%%%%%%%%%%%%%%%%%%%%%%%%%%%%%%%%%%%%%%%%%%%%%%%%%%%%

\begin{figure}[htb]
\begin{center}
\vspace{-0cm}
\leavevmode \epsfxsize=80mm 
\epsffile{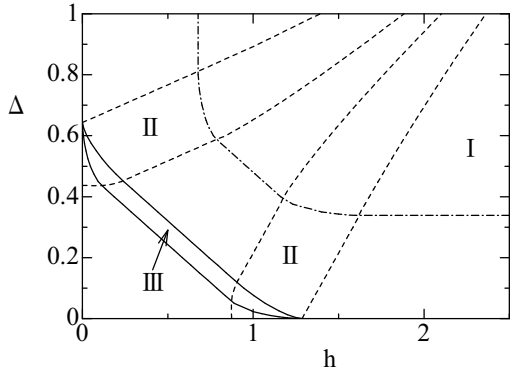}
\vspace{-6cm}
\end{center}
\caption{The $h-\Delta$ phase diagram 
for Eqs. (\protect{\ref{XXZ}}) and (\protect{\ref{crystal}}) 
at $\Phi=1.5$.
The meaning of massless regions I, II, III is same as in 
Fig.\protect{\ref{h-phi}}.
In the areas outside of these regions, the system is
in a spin-gap phase.
}
\label{h-delta}
\end{figure}

In Fig. \ref{h-delta},  we show the $h-\Delta$ phase diagram 
derived for the anisotropy parameter $\Phi=1.5$. 
Note that the phase diagram has two different regions 
labeled by II. This is because both the magnetic field and the 
crystal field  play a similar role for the formation of 
the spin gap.
%%%%%%%%%%%%%%%%%%%%%%%%%%%%%%%%%%%%%%%%%%%%%%%%%%%%%%%%%%%%%%%%%%%
%%%%%%%%%%%%%%%%%%%%%%%%%%%%%%%%%%%%%%%%%%%%%%%%%%%%%%%%%%%%%%%%%%%
 The areas outside of  the regions I, II 
and III are all massive with the spin gap, as mentioned
above.  The mechanism for the formation of plateaus and cusps 
is as discussed before.

As summarized above, the multicomponent spin system with the
spin gap exhibits interesting features such as 
plateaus and cusps in the magnetization curve,
 depending on the anisotropy parameter. 

%%%%%%%%%%%%%%%%%%%%%%%%%%%%%%%%%%%%%%%%%%%%%%%%%%%%%%%%%%%%%%%%%%%
%%%%%%%%% 3.　 NMR      　　　　　　  %%%%%%%%%%%%%%%%%%%%%%%%%%%%%
%%%%%%%%%%%%%%%%%%%%%%%%%%%%%%%%%%%%%%%%%%%%%%%%%%%%%%%%%%%%%%%%%%%
\section{NMR relaxation rate}

We have so far treated  the static magnetization, 
and have shown that the quantum phase transitions are
induced by the magnetic field, which are accompanied by the
plateau and cusp singularities. It is then interesting to 
observe how dynamical quantities behave around such
field-induced quantum phase transitions.
To address this problem, we here study the NMR relaxation rate $1/T_1$ 
in magnetic fields with the emphasis on the 
behavior around plateaus and cusps. 

The NMR relaxation rate for the multicomponent
Tomonaga-Luttinger liquid is given at low temperatures by 
\cite{schulz,sachdev1,chitra,AK}
%%%%%%%%%%%%%%%%%%%%%%%%%%%%%%%%%%%%%%%%%%%%%%%%%%%%%%%%%%%%%%%%%%%%%%%%%%%%%%%
\begin{eqnarray}
     \frac{1}{T_{1}}&=&\lim_{\omega\rightarrow 0}
                         \frac{2 k_{B}T}{\hbar^{2}\omega}
                          \int\frac{{\rm d}k}{2\pi}
                           A^{2}(k){\rm Im}\chi(k,\omega)
\nonumber \\
                    &\sim&
                       \sum_{\alpha}B_{\alpha}
                        \Gamma_{\alpha}(h)
                         T^{\eta_{\alpha}-1},
\label{nmr-sec3}
\end{eqnarray}
%%%%%%%%%%%%%%%%%%%%%%%%%%%%%%%%%%%%%%%%%%%%%%%%%%%%%%%%%%%%%%%%%%%%%%%%%%%%%%%
where 
%%%%%%%%%%%%%%%%%%%%%%%%%%%%%%%%%%%%%%%%%%%%%%%%%%%%%%%%%%%%%%%%%%%%%%%%%%%%%%%
\begin{eqnarray}
     \Gamma_{\alpha}(h)=
       \left[
             \prod_\mu 
              \left( \frac{1}{\tilde{v}_{\mu}} \right)
               ^{2x_{\alpha\mu}}
      \right],
\label{nmr2-sec3}
\end{eqnarray}
%%%%%%%%%%%%%%%%%%%%%%%%%%%%%%%%%%%%%%%%%%%%%%%%%%%%%%%%%%%%%%%%%%%%%%%%%%%%%%%
and $B_{\alpha}$ is a constant related to the hyperfine 
coupling, which is assumed to be independent of
 the temperature and the magnetic field. 
The above power-law dependence in temperature is typical for
correlation functions in the  Tomonaga-Luttinger liquid.\cite{Haldane}
Since we are now dealing with the multicomponent system,
there are several kinds of relaxation processes, 
providing different critical exponents $\eta_\alpha$, where 
the index $\alpha$ classifies elementary excitations, 
e.g. $2k_F$-current
excitation, etc.  To determine $\eta_\alpha$, we should
further incorporate the interference between different 
excitation modes,\cite{AK} which are explicitly expressed as
$\eta_{\alpha}=\sum_{\mu}2x_{\alpha\mu}$, where $x_{\alpha\mu}$
is the scaling dimension originating from the interference between 
$\alpha$ and $\mu$ modes.  This formula can be straightforwardly
obtained by analyzing the excitation spectrum
in terms of finite-size scaling techniques in conformal field 
theory.\cite{U1,Dressed,Woy,Frahm,Kawakami-1}

In the above formula, 
we have explicitly written down the factor (\ref{nmr2-sec3}) 
depending on the renormalized velocities $\tilde{v}_{\mu}$,
 which have been usually neglected in the discussion of the 
temperature dependence of $1/T_1$.  However, in order to 
study the field-dependence of $1/T_1$, we should take into
account this factor, since it plays a crucial role around the
critical points, as will be shown momentarily.
We will refer to this factor as the enhancement factor
in the following discussions.

Note that if the NMR is done on the nucleus of lattice spin,
the relaxation occurs through a contact interaction
and thereby $1/T_{1}$ depends only on the 
transverse susceptibility $\chi_{\perp}$.
On the other hand, if it is done on other neighboring
nuclei, the relaxation is through dipolar interactions, 
and thus $1/T_{1}$ depends not only on 
$\chi_{\perp}$ but also on the longitudinal
susceptibility  $\chi_{\parallel}$.  In order to
treat generic cases, we take into account 
the contributions both from  $\chi_{\perp}$ and  $\chi_{\parallel}$.
In the following, we  discuss the  critical 
properties of $1/T_1$  around plateaus and  cusps.

%%%%%%%%%%%%%%%%%%%%%%%%%%%%%%%%%%%%%%%%%%%%%%%%%%%%%%%%%%%%%%%%%%%%%%%%%%%%
%%%%%%%%%%%%%%%%%%%%　　Around Plateau　　　　 %%%%%%%%%%%%%%%%%%%%%%%%%%%%%
%%%%%%%%%%%%%%%%%%%%%%%%%%%%%%%%%%%%%%%%%%%%%%%%%%%%%%%%%%%%%%%%%%%%%%%%%%%%

\subsection{Around the plateau}

We start our discussion with the massless regions 
close to the magnetization plateau.  For instance, we 
look at the critical region beside the $m=1/3$ plateau in Fig. \ref{mag}(a), 
where only one massless mode exists both in the left-hand side (III) and the
right-hand side (II).  In the region III,  the scaling dimension $x$
is obtained  in a standard form via finite-size scaling analysis,\cite{U1} 
%%%%%%%%%%%%%%%%%%%%%%%%%%%%%%%%%%%%%%%%%%%%%%%%%%%%%%%%%%%%%%%%%%%%%%%%%%%%
\begin{eqnarray}
  x=\frac{1}{4\xi_{3}^{2}}\Delta M^2
              +\xi_{3}^{2}\Delta D^2,
\end{eqnarray}
%%%%%%%%%%%%%%%%%%%%%%%%%%%%%%%%%%%%%%%%%%%%%%%%%%%%%%%%%%%%%%%%%%%%%%%%%%%%
where so-called dressed charge\cite{Dressed} 
$\xi_3=\lim_{\lambda_3\rightarrow\lambda_3^0} 
\xi_{3}(\lambda_{3})$ is determined by
%%%%%%%%%%%%%%%%%%%%%%%%%%%%%%%%%%%%%%%%%%%%%%%%%%%%%%%%%%%%%%%%%%%%%%%%%%%%
\begin{eqnarray}
\xi_{3}(\lambda_{3}) &=&  1
          -\int_{-\lambda_{3}^{0}}
               ^{+\lambda_{3}^{0}}
               \frac{{\rm d}\lambda'_3}{2\pi}
               K_{2}(\lambda_{3}-\lambda'_{3})
                \xi_{3}(\lambda'_{3}).
\label{XXZ-d-charge1}
\end{eqnarray}
%%%%%%%%%%%%%%%%%%%%%%%%%%%%%%%%%%%%%%%%%%%%%%%%%%%%%%%%%%%%%%%%%%%%%%%%%%%%
Here $\Delta M$ ($\Delta D$) is the quantum 
number which changes the number of spinons (carries the 
$2k_F$ current).
For the longitudinal spin susceptibility $\chi_{\parallel}$ 
%%%%%$\langle S^zS^z \rangle$, 
the most relevant critical exponent is given 
by $\eta_{\parallel}=2x=2\xi_3^2$ 
by choosing  $\Delta M=0$ and $\Delta D=1$, while for  
 the transverse spin susceptibility $\chi_{\perp}$, it is given by 
%%%%%% $\langle S^+S^- \rangle$, 
$\eta_{\perp}=\frac{1}{2}\xi_3^{-2}$ 
for $\Delta M=1$ and $\Delta D=0$.
The critical exponents in region II
can be obtained via a similar analysis.

%%%%%%%%%%%%%%%%%%%%%%%%%%%%%%%%%%%%%%%%%%%%%%%%%%%%%%%%%%%%%%%%%%%
\begin{figure}[t]
\begin{center}
\vspace{-0cm}
\leavevmode \epsfxsize=80mm 
\epsffile{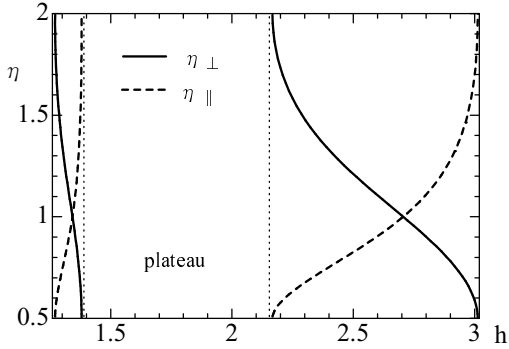}
\vspace{-6.2cm}
\end{center}
\caption{Critical exponents around  the $m=1/3$ plateau 
in Fig. \protect{\ref{mag}}(a). 
The solid lines denote the most relevant critical exponents for 
the transverse spin susceptibility while the broken lines  
that for the longitudinal susceptibility. 
}
\label{eta-plateau}
\end{figure}
%%%%%%%%%%%%%%%%%%%%%%%%%%%%%%%%%%%%%%%%%%%%%%%%%%%%%%%%%%%%%%%%%%%

\begin{figure}[b]
\begin{center}
\vspace{-0cm}
\leavevmode \epsfxsize=80mm 
\epsffile{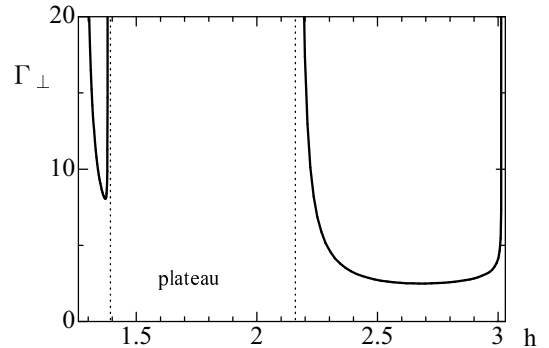}
\vspace{-6.2cm}
\end{center}
\caption{The enhancement factor $\Gamma_{\perp}$ around the plateau,
$m=1/3$, in the magnetization curve.
}
\label{gam-plateau}
\end{figure}

In Fig. \ref{eta-plateau}, we show the computed critical exponents 
as a function of the magnetic field $h$. Note that 
when the critical exponents are smaller (larger) than unity, 
 $1/T_1$ diverges (vanishes) 
with power-law dependence 
as the temperature decreases (see (\ref{nmr-sec3})). 
Similar behaviors can be found for the one-component spin-gap system 
studied  by Chitra and Giamarchi.\cite{chitra}
We here pay our attention to the enhancement factor  $\Gamma_{\alpha}$ 
at sufficiently low temperatures. In Fig. \ref{gam-plateau},
we show  $\Gamma_{\perp}$ for the most relevant sector in the 
 transverse susceptibility  as a function  of the magnetic field.
We note that $\Gamma_{\parallel}$ and  $\Gamma_{\perp}$  exhibit
essentially the same critical behaviors.  
It is seen  that $1/T_1$ is considerably enhanced in the vicinity of
 the plateau according to the dramatic renormalization of the 
velocities in the vicinity of the spin-gap phase.
The enhancement of the relaxation rate is observed whenever
the system is in a massless phase close to the spin-gap phase. 
For example, the relaxation rate  is enhanced  when 
the magnetization decreases to zero before forming the zero-field
spin gap.

%%%%%%%%%%%%%%%%%%%%%%%%%%%%%%%%%%%%%%%%%%%%%%%%%%%%%%%%%%%%%%%%%%%
%%%%%%%%%%%%%%%%%%%%%%%%%%%%%%%%%%%%%%%%%%%%%%%%%%%%%%%%%%%%%%%%%%%

%%%%%%%%%%%%%%%%%%%%%%%%%%%%%%%%%%%%%%%%%%%%%%%%%%%%%%%%%%%%%%%%%%%%%%%%%%%%
%%%%%%%%%%%%%%%%%%%%　　Around Cusp　　　　　 %%%%%%%%%%%%%%%%%%%%%%%%%%%%%
%%%%%%%%%%%%%%%%%%%%%%%%%%%%%%%%%%%%%%%%%%%%%%%%%%%%%%%%%%%%%%%%%%%%%%%%%%%%

\subsection{Around the cusp}

Let us now observe  what happens for $1/T_1$ around the cusp structure
in the magnetization.
We first look at the cusp at $h \simeq 2.1$ in Fig. \ref{mag}(b). 
The critical behavior in the left-hand side of the cusp 
is described in terms of a massless mode. Therefore, 
this region can be analyzed following the
way done for the plateau case.  We thus have the enhancement 
of $1/T_1$ near the cusp.  On the other hand, on the 
right-hand  side of  the cusp, two massless modes appear 
which need a different treatment for their 
critical properties.   According to the conformal field 
theory analysis
of multicomponent cases, the scaling dimension is 
given by the matrix form,\cite{Dressed,Woy,Frahm,Kawakami-1}
%%%%%%%%%%%%%%%%%%%%%%%%%%%%%%%%%%%%%%%%%%%%%%%%%%%%%%%%%%%%%%%%%%%%%%%%%%%%
\begin{eqnarray}
  x=\frac{1}{4}\Delta \mbox{\boldmath$M$}^T
              (\mbox{\boldmath$\xi$}^{-1})^T(\mbox{\boldmath$\xi$}^{-1})
              \Delta \mbox{\boldmath$M$}
           +\Delta \mbox{\boldmath$D$}^T
              \mbox{\boldmath$\xi$}^T\mbox{\boldmath$\xi$}
              \Delta \mbox{\boldmath$D$},
\end{eqnarray}
%%%%%%%%%%%%%%%%%%%%%%%%%%%%%%%%%%%%%%%%%%%%%%%%%%%%%%%%%%%%%%%%%%%%%%%%%%%%
%
where the quantum numbers in the vector representation are
$\Delta \mbox{\boldmath$M$}=(\Delta M_1, \Delta M_2)^T$ 
and $\Delta \mbox{\boldmath$D$}=(\Delta D_1, \Delta D_2)^T$, which 
are subject to the  selection rules,
%%%%%%%%%%%%%%%%%%%%%%%%%%%%%%%%%%%%%%%%%%%%%%%%%%%%%%%%%%%%%%%%%%%%%%%%%%%%
\begin{eqnarray}
\Delta D_1=\frac{1}{2}\Delta M_2 \;\; {\rm mod} \; 1, \;
\Delta D_2=\frac{1}{2}\Delta M_1 \;\; {\rm mod} \; 1.
\end{eqnarray}
%%%%%%%%%%%%%%%%%%%%%%%%%%%%%%%%%%%%%%%%%%%%%%%%%%%%%%%%%%%%%%%%%%%%%%%%%%%%
Here the $2 \times 2$ dressed charge matrix\cite{Dressed} 
$\boldmath \xi$ has the elements
%%%%%%%%%%%%%%%%%%%%%%%%%%%%%%%%%%%%%%%%%%%%%%%%%%%%%%%%%%%
%\begin{equation}
%\mbox{\boldmath$\xi$}=
%      \left[
%\begin{array}{cc}
%  \xi_{11}  &  \xi_{12}  \\
%  \xi_{21}  &  \xi_{22}  
%\end{array}
%      \right]
%\end{equation}
%%%%%%%%%%%%%%%%%%%%%%%%%%%%%%%%%%%%%%%%%%%%%%%%%%%%%%%%%%%%%
$\xi_{\alpha\beta}=\xi_{\alpha\beta}(\lambda_{\beta}^0)$
which are determined by
%%%%%%%%%%%%%%%%%%%%%%%%%%%%%%%%%%%%%%%%%%%%%%%%%%%%%%%%%%%%%
\begin{eqnarray}
\xi_{\alpha\beta}(\lambda_{\beta}) &=&  \delta_{\alpha\beta}
        \cr
        &&
          +
            \sum_{\gamma=1,2}
            \int_{-\lambda_{\gamma}^{0}}
               ^{+\lambda_{\gamma}^{0}}
           \frac{{\rm d}\lambda'_{\gamma}}{2\pi}
                       K_{\beta\gamma}
                           (\lambda_{\beta}-\lambda'_{\gamma})
                \xi_{\alpha\gamma}
                (\lambda'_{\gamma}),
\label{XXZ-d-charge2}
\end{eqnarray}
%%%%%%%%%%%%%%%%%%%%%%%%%%%%%%%%%%%%%%%%%%%%%%%%%%%%%%%%%%%%%%%%%%%%
where $\alpha,\beta,\gamma=1,2$,  and 
$K_{11}(\lambda)=K_{22}(\lambda)=-K_{2}(\lambda)$,
$K_{12}(\lambda)=K_{21}(\lambda)=K_{1}(\lambda)$.

\begin{figure}[t]
\begin{center}
\vspace{-0cm}
\leavevmode \epsfxsize=80mm 
\epsffile{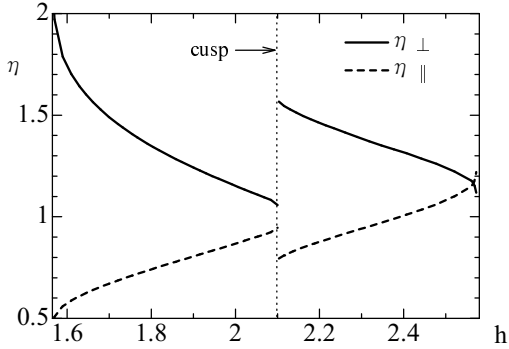}
\vspace{-6.2cm}
\end{center}
\caption{The critical exponents around the cusp. 
As an example, we show them for the cusp of $h  \simeq 2.1$ 
in Fig. \protect{\ref{mag}}(b). 
}
\label{eta-cusp}
\end{figure}

\begin{figure}[b]
\begin{center}
\vspace{-0cm}
\leavevmode \epsfxsize=80mm 
\epsffile{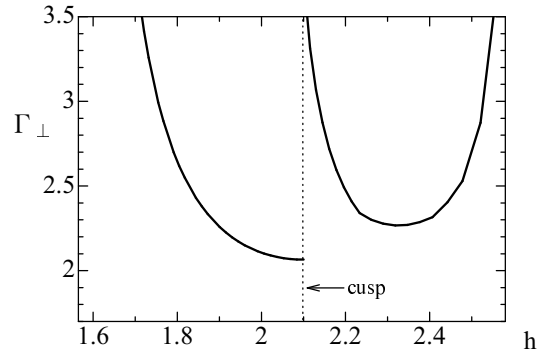}
\vspace{-6.2cm}
\end{center}
\caption{The enhancement factor $\Gamma_{\perp}$ 
for $\chi_{\perp}$ around the cusp 
in Fig. \protect{\ref{mag}}(b).
}
\label{gam-cusp}
\end{figure}

According to ref.\cite{Frahm}, by properly choosing
the set of quantum numbers $(\Delta M_1,\Delta M_2,\Delta D_1,\Delta D_2)$,
we can determine the most  relevant  critical exponents.
%%%For $\langle S^zS^z \rangle$,  
For $\chi_{\parallel}$,  we set
 $(0,0,-1,0)$, $(0,0,-1,-1)$ and $(0,0,0,-1)$, because
there is no change in the $z$-component of spin 
($\Delta M_1=\Delta M_2=0$), while for $\chi_{\perp}$,
%%%$\langle S^+S^- \rangle$, 
we set $(-1,0,0,+\frac{1}{2})$, 
$(-1,-1,-\frac{1}{2},+\frac{1}{2})$ and $(0,-1,-\frac{1}{2},0)$. 
%%%Since the choice of $(0,0,-1,-1)$ and $(0,-1,-\frac{1}{2},0)$ are 
%%similar to that for left-side of the cusp, 
In Fig. \ref{eta-cusp}.
the most relevant critical exponents obtained for 
$\chi_{\perp}$ and  $\chi_{\parallel}$
are shown as a function of the magnetic field.  It is seen that
the critical exponent $\eta_{\parallel}$ is 
smaller than $\eta_{\perp}$  in a wide range of magnetic 
fields, which  implies that $\chi_{\parallel}$ is dominant 
for the relaxation.
 As mentioned before, when the dipolar interactions are weak, 
$A_{\parallel}$ should be nearly zero, so that the divergent behavior 
in  $1/T_1$ may be masked at low temperatures in that case. 
On the other hand,  the divergent behavior should
be always observed around the plateaus, at least 
either on the right or left region. 
Fig. \ref{gam-cusp} shows the results for 
the enhancement factor $\Gamma_{\perp}$ around
the cusp in the magnetization.  In contrast to the 
case of the plateau, the enhancement of  $1/T_1$ appears
on either side of the cusp, because $\Gamma_{\alpha}$
is enhanced  when one of the massless modes becomes massive
as the magnetic field is changed.  For example, 
in the case of $\Phi=0$ in Fig. \ref{h-phi}, 
we have several quantum phase transition points at $h\simeq 0.54,1.4,6.0$,
for which we can observe the enhancement of $1/T_1$ in 
the left-hand side of these points. 

%%%%%%%%%%%%%%%%%%%%%%%%%%%%%%%%%%%%%%%%%%%%%%%%%%%%%%%%%%%%%%%%%%%
%%%%%%%%%%%%%%%%%%%%%%%%%%%%%%%%%%%%%%%%%%%%%%%%%%%%%%%%%%%%%%%%%%%
%%%%%%%%%%%%%%%%%%%%%%%%%%%%%%%%%%%%%%%%%%%%%%%%%%%%%%%%%%%%%%%%%%%
%%%%%%%%%%%%%%%%%%%%%%%%%%%%%%%%%%%%%%%%%%%%%%%%%%%%%%%%%%%%%%%%%%%

We have so far mentioned the enhancement of $1/T_1$  
around  plateaus and cusps in the magnetization. 
It is noted here that such  enhancement of $1/T_1$ is induced
not only by the magnetic field but also 
by the orbital splitting.  For instance, 
in Fig. \ref{h-delta},  the relaxation rate for $h=0$  
is enhanced near the both edges of region II,
when the orbital splitting $\Delta$ is continuously changed.

%%%%%%%%%%%%%%%%%%%%%%%%%%%%%%%%%%%%%%%%%%%%%%%%%%%%%%%%%%%%%%%%%%%
%%%%%%%%% 4. 　 SUMMARY 　　　　　　  %%%%%%%%%%%%%%%%%%%%%%%%%%%%%
%%%%%%%%%　　　　　　　　　　　　　　 %%%%%%%%%%%%%%%%%%%%%%%%%%%%%
%%%%%%%%%%%%%%%%%%%%%%%%%%%%%%%%%%%%%%%%%%%%%%%%%%%%%%%%%%%%%%%%%%%
\section{SUMMARY}

We have studied the magnetization process as well as
the spin dynamics through the NMR relaxation rate
for the 1D multicomponent spin-gap system. 
By using the solvable model of Bariev {\it et al}, 
we have found that the plateaus 
and the cusps are generated  in the  magnetization curve.  
We have also studied the NMR relaxation rate $1/T_1$ 
near  plateaus and  cusps in the magnetization.
For the critical region around the plateau, 
 $1/T_1$ diverges at
 low temperatures  at least on either 
side of the plateau, whereas 
for the critical regions around the cusp, it 
 depends on the values of the hyperfine
coupling constants
whether such a  divergent behavior  appears or not.
It has also been found that the relaxation rate 
at low temperatures is enhanced 
near the critical value of the magnetic field. This is not
only the case for the multicomponent model but also for 
ordinary spin chain models in magnetic fields. 
%#The enhancement appears at sufficiently low temperatures,
%#so that it may be observed in rather ideal 1D systems for which
%#the long-range order due to 3D dimensionality is 
%#suppressed even at low temperatures.
The enhancement appears at sufficiently low temperatures, 
so that it may be observed in rather ideal 1D spin systems for which 
the long-range order due to 3D dimensionality is suppressed even 
at low temperatures. Also, to extract this enhancement effect from 
the experimental data, careful analysis of the NMR data may be needed:
e.g. the power-law temperature dependence should be determined 
precisely to obtain the correct coefficient as a function of the 
magnetic field. Although such an enhancement has not been observed 
yet for a typical 1D spin-1/2 system CuHpCl,\cite{chaboussant} 
we hope that more detailed experiments near the critical magnetic 
field may reveal the enhancement effect discussed here. 

Although we have studied a rather specific integrable model,
some characteristic features found in this paper 
are expected to hold 
 for more generic 1D multicomponent quantum spin systems.
In this connection, 
we note here that the SU(2) $\times$ SU(2) deformation of the 
massless SU(4) model drives the system to a massive phase with
 spin gaps for all the excitation modes, according to
recent intensive studies.\cite{Pati,Azaria1,Yamashita,itoi,tsukamoto}
We think that the characteristic features obtained in
the present study can be also applied to the static and dynamical 
properties in this model.

%%%%%%%%%%%%%%%%%%%%%%%%%%%%%%%%%%
\section*{Acknowledgements}
%%%%%%%%%%%%%%%%%%%%%%%%%%%%%%%%%%%%%
This work was partly supported by a Grant-in-Aid from the Ministry 
of Education, Science, Sports and Culture of Japan. 
A. K. and F. T. were supported by 
Japan Society for the Promotion of Science. 
We would like to thank M. Sigrist for careful reading of the manuscript. 
%%%%%%%%%%%%%%%%%%%%%%%%%%%%%%%%%%%%%%%%%%%%%%%

%%%%%%%%%%%%%%%%%%%%%%%%%%%%%%%%%%%%%%%%%%%%%%%%%%%%%%%%%%%%%%%%%%%
%%%%%%%%%    　Reference  　　　　　  %%%%%%%%%%%%%%%%%%%%%%%%%%%%%
%%%%%%%%%　　　　　　　　　　　　　　 %%%%%%%%%%%%%%%%%%%%%%%%%%%%%
%%%%%%%%%%%%%%%%%%%%%%%%%%%%%%%%%%%%%%%%%%%%%%%%%%%%%%%%%%%%%%%%%%%

\clearpage
%%%%%%%%%%%%%%%%%%%%%%%%%%%%%%%%%%%%%%%%%%%%%%%%%%%%%%%%%%%%%%%%%%%

\end{document}